\pdfoutput=1
\documentclass[reprint, amsmath, amssymb, aps, prl, showpacs, superscriptaddress, groupedaddress, unsortedaddress]{revtex4-1}

\usepackage{graphicx}
\usepackage{dcolumn}
\usepackage{bm}
\usepackage{times}
\usepackage{mathrsfs}
\usepackage{color}
\usepackage{epstopdf}
\usepackage{mathrsfs}
\makeatletter
\renewcommand*{\@fnsymbol}[1]{\ensuremath{\ifcase#1\or * \else * \fi}}

\makeatother
\usepackage[colorlinks=true, letterpaper=true, pdfstartview=FitV, linkcolor=blue, citecolor=blue, urlcolor=blue]{hyperref}

\begin{document}

\title{Topological, Valleytronic, and Optical Properties of Monolayer PbS}

\author{Wenhui Wan$^\dag$}
\author{Yugui Yao$^\dag$}
\email{ygyao@bit.edu.cn}
\affiliation{$^\dag$Beijing Key Laboratory of Nanophotonics and Ultrafine Optoelectronic Systems, School of Physics, Beijing Institute of Technology, Beijing 100081, China}
\author{Liangfeng Sun$^\ddag$}
\affiliation{$^\ddag$Department of Physics and Astronomy, Bowling Green State University, Bowling Green, Ohio 43403, USA}
\author{Cheng-Cheng Liu$^\S$$^\dag$}
\author{Fan Zhang$^\S$}\email{zhang@utdallas.edu}
\affiliation{$^\S$Department of Physics, University of Texas at Dallas, Richardson, Texas 75080, USA\\
                  }              
                  
\begin{abstract}
We systematically examine the topological, valleytronic, and optical properties of
experimentally accessible PbS (001) few-layers, with a focus on the monolayer.
With even-odd layer-dependent oscillations and without band inversions,
the band gaps cover a wide spectrum from infrared to visible,
making the few-layers promising for optoelectronics.
Intriguingly, the uniaxial (biaxial) compressive strain can tune the trivial monolayer
into a $\mathcal{Z}_2$-topological (topological crystalline) insulator,
which may be utilized for controllable low-power electronic devices.
Although elliptical dichroism vanishes in the monolayer due to inversion symmetry,
optical pumping provides an efficient tool to characterize the three phases
and to realize charge, spin, and valley Hall effects in optoelectronic transport
that are tunable by external strain and light ellipticity.
\end{abstract}
\maketitle

\indent\textcolor{blue}{\em Introduction.}---
Lead sulfide (PbS)~\cite{Kohn1973} is an attractive material that has been receiving significant scientific attention.
Consisting of elements with high natural abundance, PbS can be converted into an excellent thermoelectric material.
The ZT values of PbS can even be made as high as $0.8$ at $723$~K upon nanostructuring
and enhanced to $1.1$ at $923$~K when processed with spark plasma sintering~\cite{Zhao2011}.
Due to the small effective mass ($m^*$) and the large dielectric constant ($\varepsilon$),
PbS exhibits an exciton Bohr radius ($a_0\!\sim\!m^*/\varepsilon$) as large as $20$~nm~\cite{Wise2000}.
Strong quantum confinement, a determining characteristic for a quantum dot, can thus be easily achieved in PbS.
Synthetic techniques can control the dot sizes and tune the band gaps of PbS colloidal dots from $0.7$ to $2.1$~eV,
spanning an ideal range for single- and multi-junction photovoltaic device applications~\cite{Sun2012,Aerts2014}.

Recently, with large spin-orbital couplings (SOC) and $L$-point band inversion,
IV-VI semiconductors SnTe/SnSe in the rocksalt structure have been demonstrated~\cite{SnTe2012,Tanaka2012,Dziawa2012,Xu2012}
to be 3D topological crystalline insulators (TCI) protected by mirror symmetries.
Although 3D PbS is topologically trivial,
its thin films~\cite{Liu2014, Wrasse2014,Wang2014, Feng1, Feng2, Liu2015, Kim2015, Safaei2015, Niu2015, Li2016}
are predicted to be 2D TCI's depending on the thickness~\cite{Liu2014, Wrasse2014, Liu2015, Kim2015},
and a transverse electric field can switch on/off the topological edge conducting channels~\cite{Liu2014},
based on first-principles calculations using the Perdew-Burke-Ernzerhof (PBE) functional~\cite{Perdew1996}.
Excitingly, quasi-2D nanoplates of IV-VI semiconductors~\cite{Shen2014,Safdar2015}
and nanocrystalline PbS (001) films with thickness of a few atomic layers
have recently been synthesized~\cite{Bhandari2014,Sun-new}.

Therefore, understanding the unique properties of PbS few-layers
at a microscopic view is of fundamental importance.
The preliminary PBE calculations may under- or over-estimate the band gaps,
whose signs are decisive for determining the topological nature of the few-layers.
Hence more advanced calculations are immediately called for.
Applying the transverse electric field is not practical for a flat monolayer.
In contrast, asserting external strain is more efficient
in tuning the band topology and switching on/off the edge channels.
Here, we perform first-principles calculations using the more accurate
Heyd-Scuseria-Ernzerhof (HSE) hybrid functional~\cite{Paier2006}
to examine the band structures of PbS (001) few-layers, with a focus on the monolayer.
We reveal that the band gaps of few-layers exhibit even-odd layer-dependent oscillations without band inversions.
We demonstrate that the uniaxial (biaxial) compressive strain can tune the monolayer to a 2D TI (TCI).
Although inversion symmetry dictates the elliptical dichroism to vanish,
optical pumping provides an efficient tool to characterize the three topological phases
and to facilitate tunable charge, spin, and valley Hall effects.

\begin{figure}[b!]
\centerline{\includegraphics[width=0.43\textwidth]{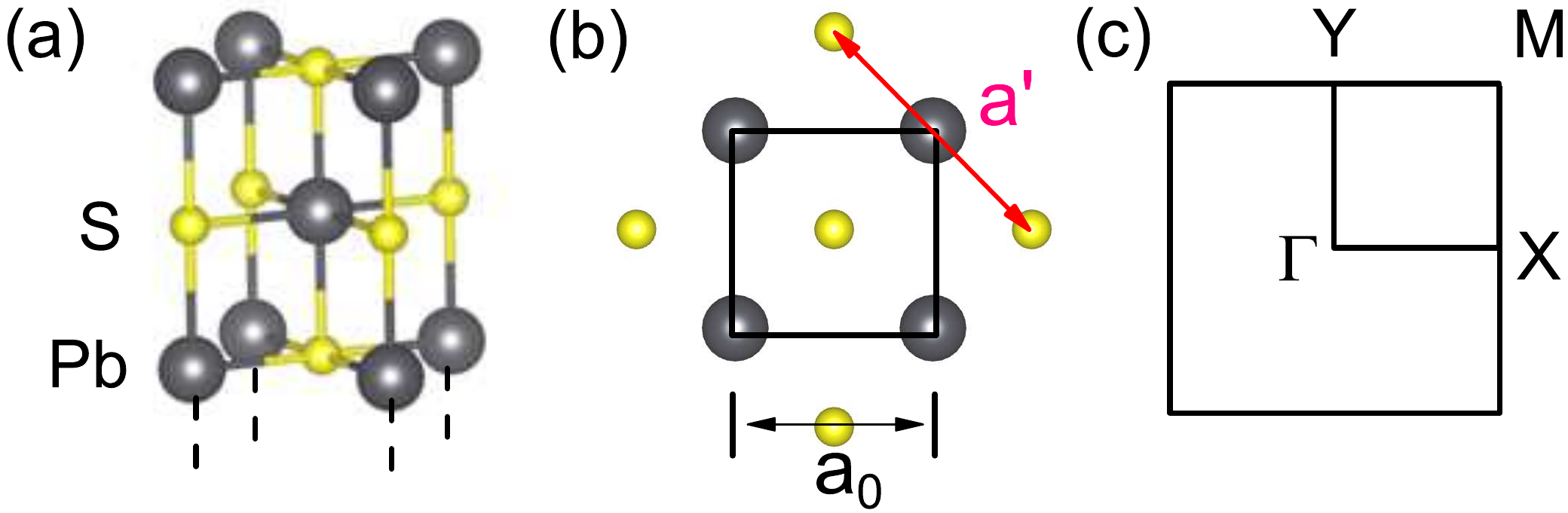}}
\caption{(a) The side view, (b) the top view, and (c) the first Brillouin zone of a PbS (001) few-layer.
In (b), $a_0$ is the few-layer lattice constant; $a'$ is the bulk lattice constant.}
\label{wh1}
\end{figure}

\indent\textcolor{blue}{\em Band structures.}---
The first-principles calculations are performed
using the projector augmented plane waves method~\cite{Blochl1994}
implemented in the Vienna {\it ab initio} simulation package~\cite{Kresse1993,Kresse1996}.
A Monkhorst-Pack grid~\cite{Monkhorst1976} of $10\times10\times1$, a vacuum layer of $16$~{\AA},
and a plane-wave energy cutoff of $400$~eV are used.
Both the lattice constants and the ion positions are allowed to be optimized
until the force on each ion is less than $0.01$~eV$\cdot${\AA}$^{-1}$.
The HSE hybrid functional~\cite{Paier2006},
more accurate and reliable than the PBE functional~\cite{Perdew1996},
is applied to the calculations of electronic band structures with SOC included.

\begin{figure}[t!]
\centerline{\includegraphics[width=0.43\textwidth]{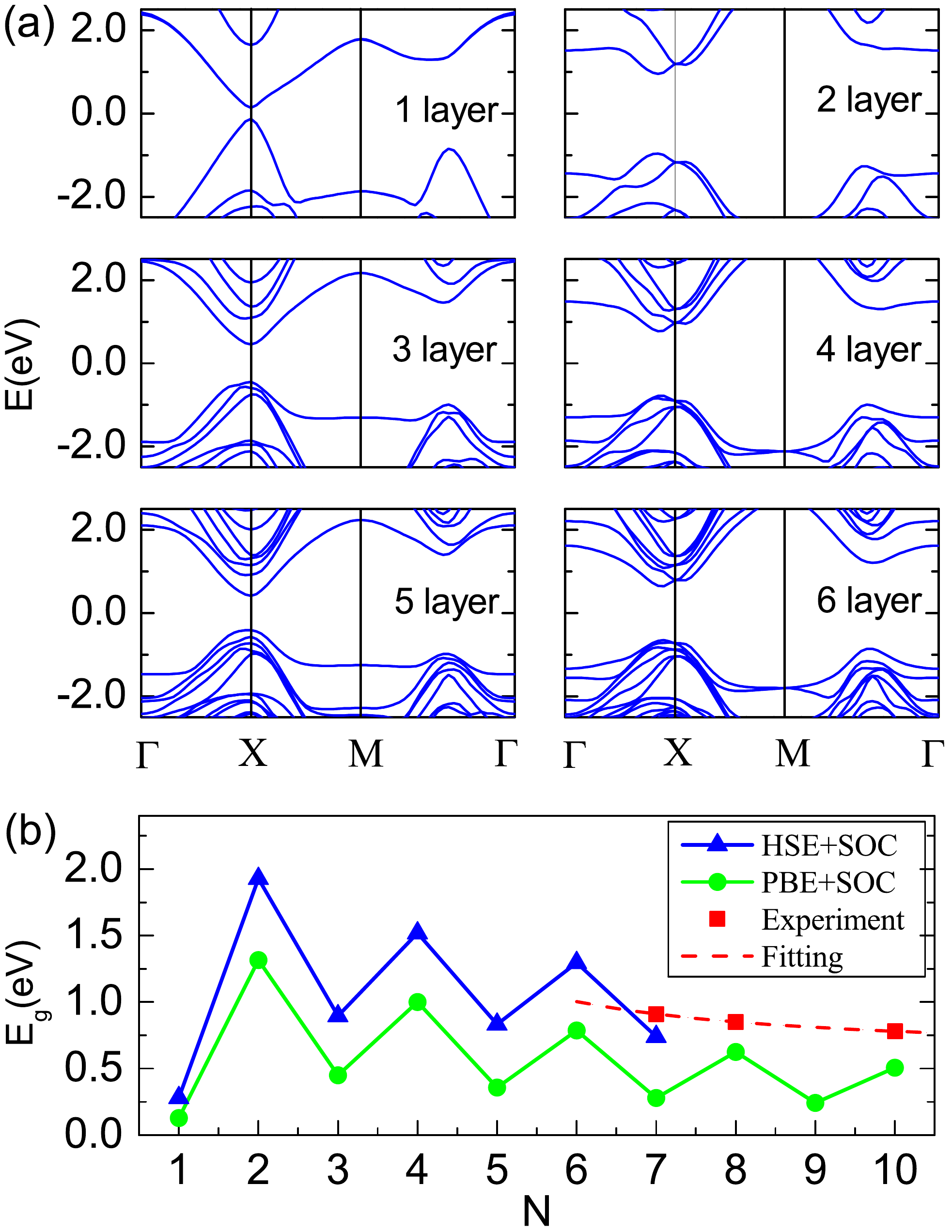}}
\caption{The band structures of PbS (001) few-layers.
(a) The band structures from monolayer to hexalayer,
obtained by the HSE+SOC method.
(b) The layer ($N$) dependence of the band gaps ($E_g$),
obtained by the PBE+SOC method, by the HSE+SOC method,
and by an optical experiment~\cite{Sun-new}. 
the experimental data.}
\label{wh2}
\end{figure}

Figure~\ref{wh1} shows the lattice structure and
the first Brillouin zone (BZ) of PbS (001) few-layers.
The calculated lattice constant of bulk PbS is $a'=6.000$~\AA,
in agreement with the experimental value of $5.936$~\AA~\cite{Madelung2005}.
The interlayer distances of the few-layers have periodic contractions and extensions.
The even-layers are made of a combination of bilayer blocks,
whereas the odd-layers each has an extra single-layer or trilayer block in its center.
The band structures of PbS (001) few-layers based on our HSE+SOC method,
from monolayer to hexalayer, are shown in Fig.~\ref{wh2}(a).
To compare the energy gaps, Fig.~\ref{wh2}(b) also plots the results from an optical experimental~\cite{Sun-new},
as well as the less accurate calculations using PBE functional. 
Clearly, the band gaps obtained by the HSE+SOC method
are closer to the experimental values. 
While the direct band gap gradually increases to the bulk value as the layer number increases,
it oscillates strongly~\cite{Liu2010} between odd- and even-layers.
A similar phenomenon has also been observed in the PbSe quantum wells~\cite{Allan2004}.
Our calculations reveal that in even-layers the relatively smaller interlayer distances within bilayer blocks
result in stronger interlayer orbital hybridization and thus their larger band gaps.
Moreover, we find that none of these few-layers exhibit any band inversion,
revising previous results~\cite{Liu2014, Wrasse2014, Liu2015}.
Intriguingly, the few-layer band gaps vary from $0.24$ to $1.92$~eV,
covering a wide spectrum from infrared to visible.
Such a property makes PbS (001) few-layers promising for optoelectronic applications.

\begin{figure}[b!]
\centerline{\includegraphics[width=0.43\textwidth]{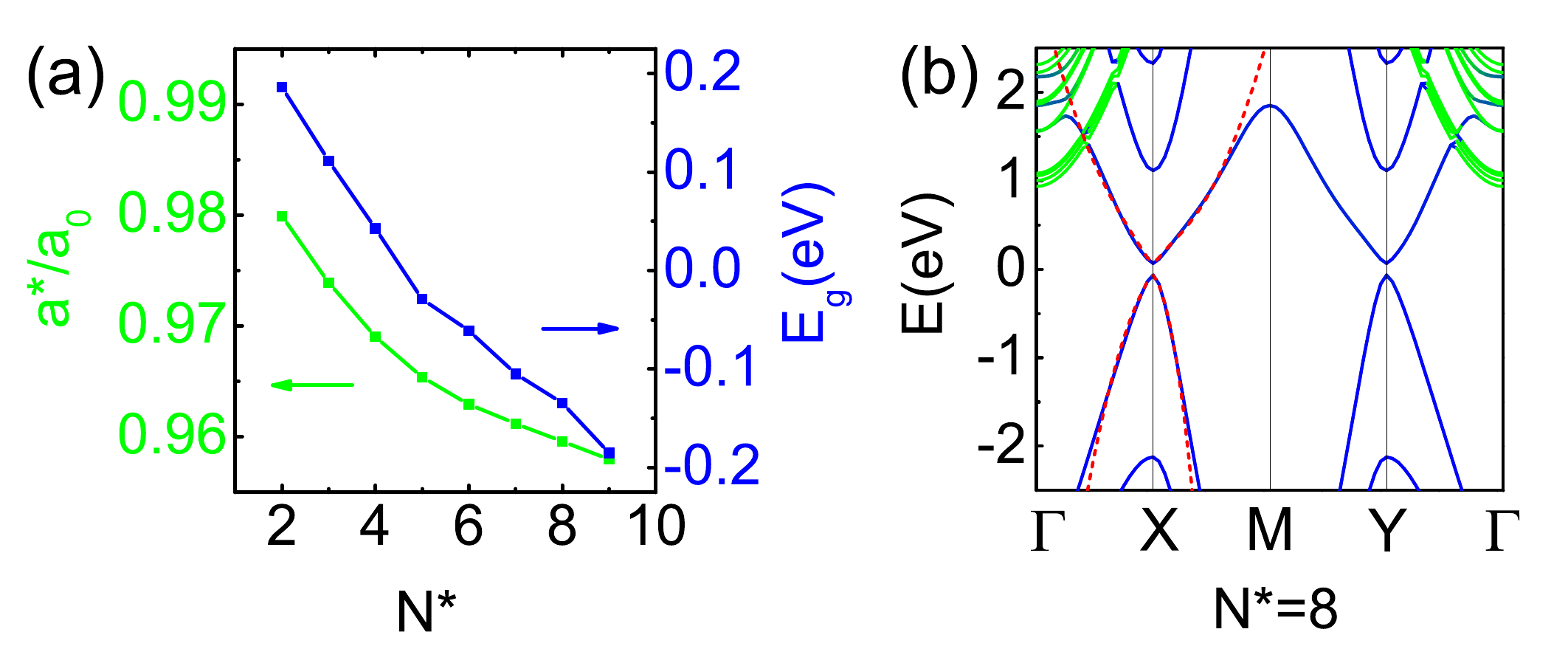}}
\caption{The lattice constants and the band structures for the commensurate KF-PbS monolayer-KF.
(a) The lattice constant ($a^{\ast}$) and the band gap ($E_{g}$), as functions of the KF layer number ($N^{\ast}$).
$a_{0}$ is the lattice constant of the flat freestanding monolayer.
(b) The band structure with $N^{\ast}=8$.
The blue and green colors indicate the contributions from PbS and KF, respectively.
The red lines are the fitted bands using the effective $k\cdot p$ Hamiltonian Eq.~(\ref{heff}).}
\label{wh3}
\end{figure}

Unexpectedly, the monolayer PbS has a sharply reduced band gap.
This arises for two reasons.
First, the bulk crystal has cubic symmetries,
and a strong crystal field effect is present in the monolayer~\cite{Liu2015}.
Second, the small band gap is also consistent with the fact
that a pressure can decrease the band gap of PbS~\cite{Barone2013};
The lattice constant of monolayer is $a_{0}\!=\!4.069$~{\AA},
equivalently $a'\!=\!\sqrt{2}a_{0}\!=\!5.754$~{\AA} in Fig.~\ref{wh1}(b),
which is smaller than the aforementioned bulk value $a'\!=\!6.000$~{\AA}.
Furthermore, we find that a flat monolayer is unstable toward
buckling Pb and S sublattices in opposite out-of-plane directions.
In experiment, PbS few-layers can be stabilized by sandwiching them in between CdS shells~\cite{Lechner2014,Sun-new}.
We hereby propose an improved structure after a systematic examination,
i.e., the monolayer sandwiched in between two KF layers.
To avoid the unbearable computational cost of the HSE+SOC calculations,
we adopt the PBE+SOC method to show the band structure of KF-PbS monolayer-KF.
As shown in Fig.~\ref{wh3}(a), the strain on the PbS monolayer is less than $4.5\%$
and tolerable in the synthesis~\cite{Gaiduk2008,Eic2014}.
As shown in Fig.~\ref{wh3}(b),
the electronic states near the Fermi level are mainly contributed by the PbS monolayer,
while the contribution from the KF layers is negligible because of their large band gaps.

\indent\textcolor{blue}{\em Strain effects.}---
As evidenced in Fig.~\ref{wh3}(a),  both the lattice constant and the band gap of the sandwiched structure decrease
as the number of the KF layers increases.
Beyond the KF tetralayer, intriguingly, band inversions occur at both $X$ and $Y$ points in the 2D BZ.
Inspired by this observation, we investigate how a strain modulates the PbS monolayer band structure
and whether the resulting one is topologically nontrivial.
We define the strain as $\varepsilon_{i}\!=\!(a_i-a_0)/a_0$ along the $i$ direction,
where $a_i$ and $a_0$ are the lattice constants of PbS monolayer with and without strain, respectively.
When $\varepsilon_{x,y}=0$, our calculations reveal that the conduction band minima (CBM)
are dominated by the $p_z$ orbital (odd parity) of Pb at both $X$ and $Y$ points,
whereas the valence band maxima (VBM) are respectively dominated by the $p_x$ and $p_y$ orbitals of S (even parity),
hybridized with the $s$ orbital of Pb, at $X$ and $Y$ points.
Importantly, this indicates~\cite{Fu-Kane} a normal insulator (NI) without band inversions.

\begin{figure}[t!]
\centerline{\includegraphics[width=0.43\textwidth]{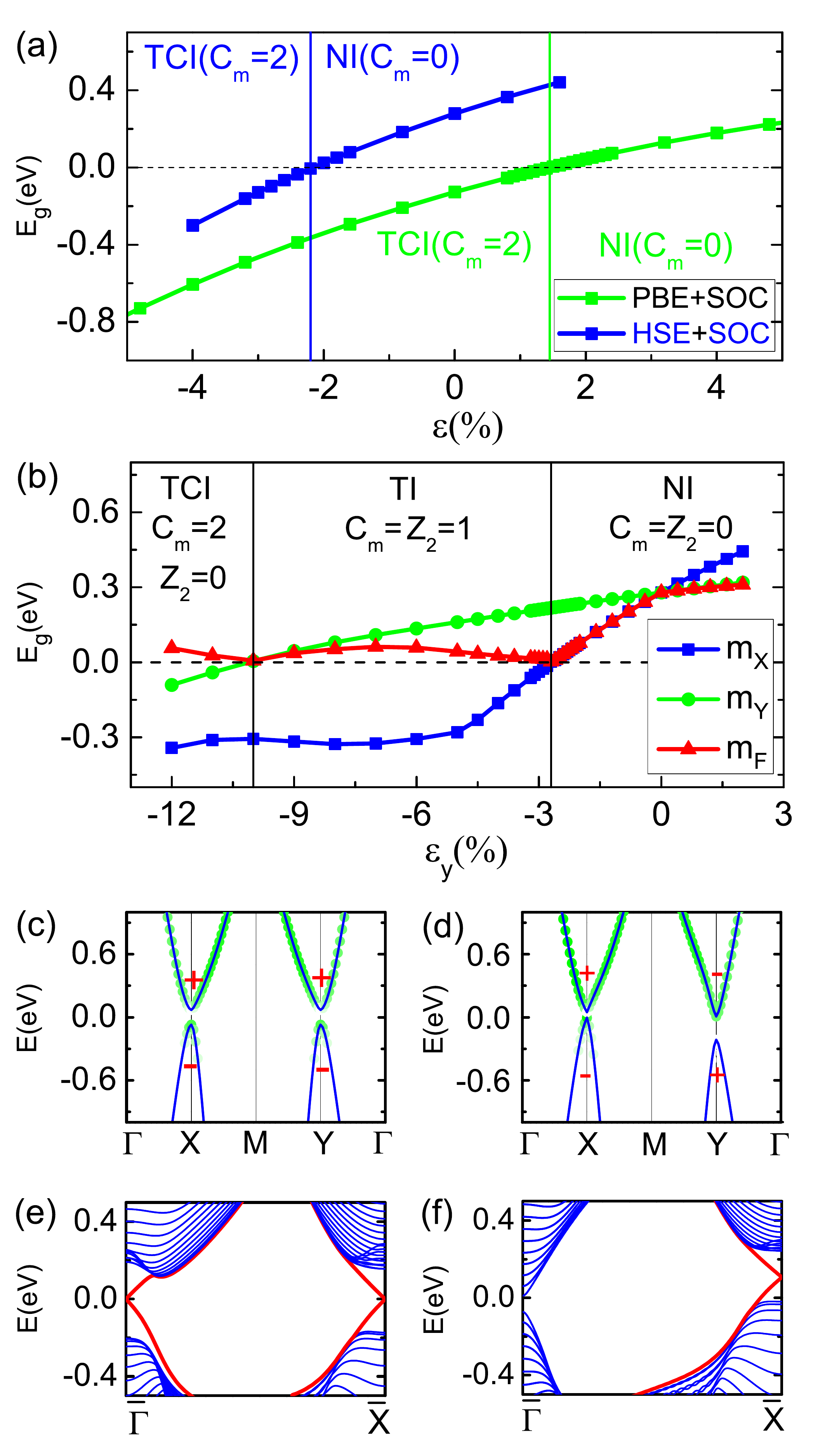}}
\caption{Strain-induced topological phases in the PbS (001) monolayer.
(a) The band gaps ($E_{g}$) and the phase diagrams as functions of the biaxial strain ($\varepsilon$),
calculated by HSE+SOC and PBE+SOC methods.
(b) The band gaps at $X$ and $Y$ points ($m_{X,Y}$), the fundamental gap ($m_{F}$) across the entire 2D BZ, and the phase diagram
as functions of the uniaxial strain ($\varepsilon_{y}$), calculated by the HSE+SOC method.
(c) The even number of band inversions induced by a biaxial strain.
(d) The odd number of band inversions induced by a uniaxial strain.
In (c) and (d), the green color represents the component of Pb-$p_{z}$ orbital;
the $\pm$ denotes the parity eigenvalues of the states at VBM and CBM.
(e)-(f) The corresponding band structures of 17 nm nanoribbons corresponding to (c) and (d), respectively.
In (e) and (f), the red lines are the edge states, and the strains larger than those in (c) and (d) are used for clarity.}
\label{wh4}
\end{figure}

We now consider the biaxial strain effects based on the HSE+SOC method.
Because of the intact $\mathcal{C}_4$ symmetry,
the parity eigenvalues of the energy bands must be the same at $X$ and $Y$ points.
This fact prevents the PbS monolayer from turning into a $\mathcal{Z}_2$ TI~\cite{Kane-Mele} under a biaxial strain.
However, a strained monolayer can be a 2D TCI, as we now demonstrate.
Our calculations reveal that a biaxial compressive strain ($\varepsilon_{x}\!=\!\varepsilon_{y}$)
stronger than $-2.2\%$ can produce band inversions at the $X$ and $Y$ points in the 2D BZ,
as shown in Figs.~\ref{wh4}(a) and~\ref{wh4}(c).
Evidently, a flat monolayer respects a mirror symmetry ($z\rightarrow -z$).
It follows that there exists a mirror Chern number $\mathcal{C}_m$~\cite{SnTe2012}.
We find that $\mathcal{C}_m=0$ for any tensile strain or a compressive strain weaker than $-2.2\%$
whereas $\mathcal{C}_m=2$ for the case with a stronger compressive strain, as summarized in Fig.~\ref{wh4}(a).

To visualize the bulk-boundary correspondence dictated by $\mathcal{C}_M=2$,
we calculate the band structure of a strained $17$~nm wide nanoribbon (a TCI) terminated by S atoms.
As clearly seen in Fig.~\ref{wh4}(e), there are two pairs of states cross the band gap,
with Dirac-like crossing points at $\bar{X}$ and $\bar{\Gamma}$ points.
Our real-space charge distribution calculations further confirm that these states
are indeed the anticipated edge states protected by the mirror symmetry.
As reflected by Fig.~\ref{wh4}(a), our finding on the unstrained PbS monolayer
based on the more advanced HSE+SOC method
is qualitatively different from previous ones~\cite{Liu2014, Wrasse2014, Liu2015}
that adopted the less accurate PBE+SOC method.
However, the phase diagrams (not considered previously) obtained by the two methods share the same qualitative trend:
the TCI phase can be achieved under a biaxial compressive strain whereas the NI gap is enlarged by a tensile strain.

We further consider the effects of a uniaxial strain $\varepsilon_{y}$ along the $y$ direction,
which breaks the $\mathcal{C}_4$ symmetry and
thus allows the energy bands at $X$ and $Y$ points to exhibit opposite parity eigenvalues.
We find that the CBMs respond little to the strain due to their Pb-$p_z$ orbital nature,
and that the energy of VBM at $X$ point rises more rapidly than that at $Y$ point, as shown in Fig.~\ref{wh4}(b).
The VBM at $X$ point has the character of an anti-bonding state between Pb-$s$ and S-$p_{x}$ orbitals,
and its energy is determined by the integral $E_{sp_{x}}^X=\langle\psi_{s}|H|\psi_{p_{x}}\rangle$,
which is proportional to the $x$-direction cosine of the vector from Pb to S atoms;
a similarly argument applies to the VBM at $Y$ point.
Since the compressive strain $\varepsilon_{y}$ decreases the $y$-direction distance between Pb and S atoms,
the band gap must be first inverted at $X$ point.
Nevertheless, when $\varepsilon_{y}$ lies between $-2.7\%$ and $-10.0 \%$,
only one band inversion occurs at $X$ point and a fundamental band gap exists across the entire 2D BZ.
When $\varepsilon_{y}$ goes beyond $-10.0 \%$, the bands at $Y$ point become inverted, too.
Based on the Fu-Kane criterion~\cite{Fu-Kane}, the former phase must be a $\mathcal{Z}_2$ TI while the latter one $\mathcal{Z}_2$ trivial.

Since the aforementioned mirror symmetry remains intact,
we can also calculate the mirror Chern numbers $\mathcal{C}_m$~\cite{SnTe2012} for the three phases.
We find that $\mathcal{C}_m$ switches from 0 to 1 and then to 2 at $\varepsilon_{y}=-2.7\%$ and $-10.0 \%$, respectively.
Furthermore, We calculate the edge states of the TI phase using the same method as we did for the TCI case.
With anisotropic strains $\varepsilon_{y}=-3.0\%$ and $\varepsilon_{x}=3.5\%$, as plotted in Fig.~\ref{wh4}(f).
the band structure of the 17 nm nanoribbon hosts only one helical edge states near $\bar{X}$ point.
For $\varepsilon_{y}$ stronger than $-10.0\%$ the band gap at $Y$ point is also inverted,
and a TCI phase with similar band structure to Fig.~\ref{wh4}(e) is identified.
Therefore, the uniaxial strain offers a controllable way to induce topological phase transitions among the NI, TI, and TCI phases,
as well as to switch the number of helical edge states among 0, 1, and 2.
We note that the TI and TCI phase exhibit dissipationless quantum spin Hall (QSH) transport via their helical edge channels.
These unique features, together with the relatively larger band gaps compared to HgTe/CdTe~\cite{HgTe} and InAs/GaSb~\cite{AsIn} QSH systems,
make the PbS monolayer promising for controllable low-power electronic devices.

\begin{figure}[t!]
\centerline{\includegraphics[width=0.9\columnwidth]{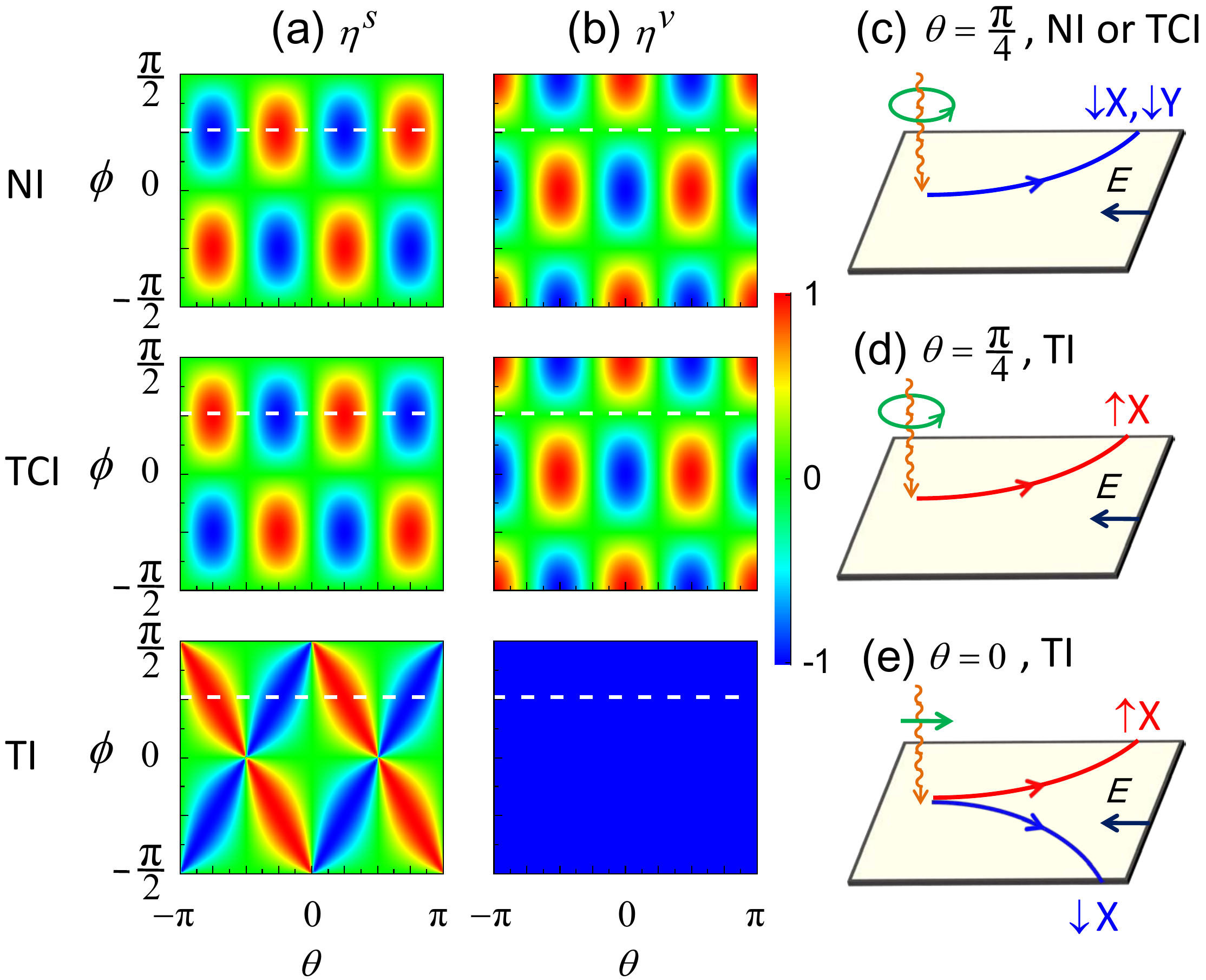}}
\caption{The optical spin-valley polarizations and the anomalous Hall effects in optoelectronic transport.
(a) The optical spin polarization $\eta^s$ and (b) the optical valley polarization $\eta^v$ for the three phases.
$\theta$ is the light ellipticity, and $\phi\!=\!\tan^{-1}(v_y/v_x)$ is the band anisotropy.
The white dashed line represents $\phi\!=\!0.26\pi$ for the PbS (001) monolayer,
obtained by fitting Fig.~\ref{wh2}(b) to Eq.(\ref{heff}).
(c)-(e) Schematic figures for the charge-, spin-, and valley-Hall effects for the photo-excited electrons in the three phases.
$\theta\!=\!0$ and $\pi/4$ denote the linearly and circularly polarized lights, respectively.}
\label{wh5}
\end{figure}

With the above knowledge, we now construct an effective model to describe the PbS (001) monolayer.
We choose the Pauli matrices $\bm \sigma$ to denote the electron spin and
$\tau_z = \pm 1$ to represent the conduction and valence bands near the band gap.
Given the little group $\mathcal{D}_{2h}$ at the $X$ point,
we further choose time reversal, spatial inversion, and three mirror reflection operators to be
$\mathcal{T}=i\sigma_y\mathcal{K}$, $\mathcal{P}=\tau_z$,  $\mathcal{M}_x=i\tau_z\sigma_x$,
$\mathcal{M}_y=i\sigma_y$, and $\mathcal{M}_z=i\sigma_z$, respectively.
To the linear order the symmetries dictate the $k\!\cdot\! p$ Hamiltonian around $X$ to be
\begin{eqnarray} \label{heff}
\mathcal{H}_{X} &=& (m_{X} + m_x k_x^2 + m_y k_y^2 ) \,\tau_z\nonumber\\
&+&(\upsilon_x k_x\tau_y + \upsilon_y k_y \tau_x \sigma_z) +(c_x k_x^2 + c_y k_y^2 )\,,
\end{eqnarray}
where $\upsilon$'s are the Fermi velocities, $m$'s are the mass terms,
and $c$'s break the electron-hole symmetry.
$\mathcal{H}_{Y}$ can be obtained by a $\mathcal{C}_4(\hat{z})$ rotation of $\mathcal{H}_{X}$ around $\Gamma$.
Essentially, the mass terms capture all the physics of band inversions.
In fact, the symmetries dictate any strain to only renormalize the $m_{X,Y}$ terms to the lowest order.
The unstrained monolayer is a NI, i.e., all $m$'s are positive.
Under a uniaxial strain stronger than $-2.7\%$ but smaller than $-10.0\%$,
the monolayer is a TI, i.e., only one of $m_{X,Y}$ switches signs.
Under a uniaxial (biaxial) strain stronger than $-10.0\%$ ($-2.2\%$),
the monolayer is a TCI, i.e., both $m_{X,Y}$ switch signs.
The red lines in Fig.~\ref{wh3}(b) plots the fitted bands using Eq.~(\ref{heff}),
and $E_g$'s in Fig.~\ref{wh4} show the $m_{X,Y}$ in various scenarios.
With $\mathcal{P}$ and $\mathcal{M}_z$ symmetries of $\mathcal{H}_{X,Y}$,
we further compute the $\mathcal{Z}_2$ and $\mathcal{C}_m$ topological invariants 
and obtain the same values as above.

\indent\textcolor{blue}{\em Optical pumping.}---
Like transition metal dichalcogenide (TMD) monolayers~\cite{TMD},
the PbS (001) monolayer also has two valleys, i.e., $X$ and $Y$.
One may naturally wonder whether the PbS (001) monolayer has a similar circular dichroism~\cite{TMD}.
In general, the elliptical dichroism is proportional to the valence-band Berry curvature.
It follows that the elliptical dichroism must vanish at each valley,
as the $\mathcal{P}$ and $\mathcal{T}$ symmetries together dictate the Berry curvature to be zero.
This result is independent of the topological nature of PbS (001) monolayer,
as a strain does not break $\mathcal{P}$ or $\mathcal{T}$ symmetry.
(In TMD monolayers, the broken $\mathcal{P}$ symmetry and the intact $\mathcal{T}$ symmetry
lead to the opposite nontrivial Berry curvatures
and hence the opposite circular dichroism at $K$ and $K'$ valleys.)

However, there are strong spin-valley polarizations in the optical absorbance of PbS (001) monolayer.
It is the $\mathcal{M}_z$ symmetry that allows the decoupling between the two spins.
The optical spin (s) and valley (v) polarizations can be defined as
\begin{flalign}\!\!\!
\eta^{s}\!=\!\frac{ \sum\limits_{\sigma,\alpha}\sigma|\mathcal{P}^{\sigma}_{\alpha}|^{2} }
{ \sum\limits_{\sigma,\alpha}|\mathcal{P}^{\sigma}_{\alpha}|^{2} },\;\qquad
\eta^{\textsl{\textrm{v}}}\!=\!\frac{ \sum\limits_{\sigma,\alpha}\alpha|\mathcal{P}^{\sigma}_{\alpha}|^{2} }
{ \sum\limits_{\sigma,\alpha}|\mathcal{P}^{\sigma}_{\alpha}|^{2} },\;
\end{flalign}
where $\sigma\!=\!\pm$ denote the up and down spins,
and $\alpha\!=\!\pm$ represent the $X$ and $Y$ valleys.
For an elliptically polarized light, $\mathcal{P}(\theta,\phi)=\mathcal{P}_{x} \cos\theta+ i \mathcal{P}_{y}\sin\theta$,
where $\mathcal{P}_{x,y}\!=\!\langle \psi_c|\hat{p}_{x,y}|\psi_v\rangle$ are the optical matrix elements,
$\theta$ is the light ellipticity, and $\phi\!=\!\tan^{-1}(\upsilon_{y}/\upsilon_{x})$ is the band anisotropy.
Focusing on the interband transitions characteristic to the $X$ and $Y$ points, we then obtain
\begin{eqnarray}\label{PX}
|\mathcal{P}_{X}^{\sigma}|^2 &=& m_{e}^{2} \upsilon^{2}\,\cos^2[\phi+\theta\,\sigma\,\textrm{sgn}(m_X)]\,,\\
|\mathcal{P}_{Y}^{\sigma}|^2 &=& m_{e}^{2} \upsilon^{2}\ \sin^2[\phi-\theta\,\sigma\,\textrm{sgn}(m_Y)].
\label{PY}
\end{eqnarray}
with $\upsilon=(\upsilon_{x}^{2}+\upsilon_{y}^{2})^\frac{1}{2}$. The combined factor $\theta\sigma$ in Eqs.~(\ref{PX})-(\ref{PY}) immediately suggests that the elliptical dichroism
$\sim\sum_{\sigma}(|\mathcal{P}^{\sigma}_{\alpha}(\theta)|^2 - |\mathcal{P}^{\sigma}_{\alpha}(-\theta)|^2)$
vanishes for each valley, consistent with our symmetry argument.
Given $m_X\!=\!m_Y$ for the NI and TCI phases,
it follows from Eqs.~(\ref{PX})-(\ref{PY}) that $\eta^{s}\!=\!-\sin(2\phi)\sin[2\theta\, \textrm{sgn}(m)]$.
Thus, the NI and TCI phases exhibit opposite spin polarizations,
which are prominent for circularly polarized lights $\theta\!=\!\pi/4$, as seen in Fig.~\ref{wh5}(a).
Similarly, we find $\eta^{\textsl{\textrm{v}}}\!=\!\cos(2\phi)\cos(2\theta)$,
as plotted in Fig.~\ref{wh5}(b);
the NI and TCI phases share the same valley polarization that is pronounced (vanishing) for linearly (circularly) polarized lights.
For the TI phase, the inverted valley has a much smaller gap,
and $\eta^{s,v}$ are dominated by the valley close to the probing threshold.
Evidenced by Figs.~\ref{wh5}(a) and~\ref{wh5}(b),
$\eta^{\textsl{\textrm{v}}}\!=\!1$ reflecting the perfect valley polarization,
and $\eta^{s}\!\sim\!\sin(2\phi)\sin(2\theta)$
indicating that the spin polarization is vanishing (pronounced) for linearly (circularly) polarized lights.

Besides characterizing the three phases,
the spin-valley selection of optical pumping may facilitate the realization
and control of intriguing Hall effects in optoelectronic transport.
Such anomalous bulk transport is due to the {\it geometric} Berry curvature
of photo-excited states at the conduction bands,
in contrast to the quantized edge transport dictated
by the {\it topological} invariant of the entire valence bands.
Based on Eq.~(\ref{heff}), we derive the Berry curvature
${\Omega}^{\sigma}_{\alpha}\!=\!{\sigma\,\textrm{sgn}(m_{\alpha})\upsilon_x\upsilon_y}\hat{z}/{2m_{\alpha}^{2}}$
for the conduction-band edge. Upon the application of an in-plane electric field ${E}\hat{x}$,
the excited electron acquires an anomalous transverse velocity
$e{E}{\Omega}^{\sigma}_{\alpha}\hat{y}$.
This implies tunable charge, spin, and valley Hall effects upon optical pumping and strain,
as illustrated in Figs.~{\ref{wh5}}(c)-{\ref{wh5}}(e).

\indent\textcolor{blue}{\em Conclusion.}---
In conclusion, we have shown that the band gaps of PbS (001) few-layers
exhibit even-odd layer-dependent oscillation without any band inversion,
by carrying out first-principles calculations employing the HSE hybrid functional.
These results should be more reliable than those perviously obtained by the less accurate PBE functional.
In particular, we reveal that the uniaxial (biaxial) compressive strain
can tune the monolayer to a 2D TI (TCI).
Hence, PbS monolayer is promising for controllable low-power electronic devices.
Although elliptical dichroism vanishes in the monolayer due to inversion symmetry,
optical pumping provides an efficient tool to characterize the three topologically different phases
and to facilitate the realization of charge, spin, and valley Hall effects
that are tunable by external strain and light ellipticity.
These unique properties, together with their band gaps covering a wide spectrum from infrared to visible,
making the PbS few-layers and particularly the monolayer as a fertile ground for
topological, valleytronic, and optoelectronic studies.
Finally, we note that similar results are anticipated for other IV-VI semiconductors.

\indent\textcolor{blue}{\em Acknowledgement.}---
WW, YY, and CCL were supported by the MOST Project of China (Nos. 2014CB920903 and 2016YFA0300603), 
the National Nature Science Fundation of China (Grant Nos. 11574029 and 11225418). 
LS was supported by the Office of the Vice President for Research \& Economic Development at Bowling Green State University.
CCL and FZ were supported by the research enhancement funds at the University of Texas at Dallas.

\bibliography{Silicene}

\end{document}